# Novel Hard Link-Switching Scheme using Pre-Scanning for Indoor VLC Networks

Sumita Mishra, Sachin Kumar,
ECE Department,
Amity University,
Lucknow Campus, India
mishra.sumita@gmail.com

Shivani Singh, Pallavi Asthana,
ECE Department,
Amity University,
Lucknow Campus, India

Nidhi Mathur
Member IETE,
Lucknow Chapter
India

*Abstract*— **VLC is attracting a lot of attention as an emerging potential technology for deployment in next generation indoor wireless networks. Use of efficient link switching scheme among VLC access points is critical in indoor environment to provide seamless connectivity to mobile users. This paper presents a novel position prediction link switching scheme for indoor visible light communication systems. The method exploits the fact that indoor scenario (light fixtures/furniture) mostly remains unchanged; therefore, this information can be stored at the coordinator end. Thus, the user is not required to search for the best transmitter when RSS is reduced to a threshold value as in conventional methods which cause substantial delay in switching process. Simulation results show that the proposed scheme for indoor VLC System achieves the dual purpose of adequate illumination and mobility to user in considered indoor scenario.**

*Keywords—Visible Light Communication; SNR; Handover; Link Switching; Received Signal Strength*

## I. INTRODUCTION

Visible light communication (VLC) is one of the promising wireless communication technology, which utilizes RGB or white LEDs to fulfill the dual requirement of communication and illumination in indoor scenario [1]. IEEE standard for Visible-light Personal Area Networks (VPANs) [2] has increased the research interest in developing and exploring techniques to achieve high data rates and reliable communication using common LEDs. Handover process of conventional wireless systems is referred as Link switching in VLC. Link switching is a critical parameter that deals with the signal degradation because of mobility of user device. Mobility can be categorized as; physical or logical Physical mobility is change of position of user device due to the movement within a cell; while logical mobility refers to change in device communication link due to interference. Link switching techniques fall into two categories namely vertical and horizontal. Vertical handover refers to the automatic switching of datalink layer technology used to access the network to maintain desired QoS. There has been lot of efforts towards developing suitable handover mechanisms in vertical handover domain. [3-6]. However, horizontal handover when switching between light sources is required due to interference or change in receiver position has not garnered much attention. T. Nguyen et al presented a novel RSS (Received signal Strength) prediction technique to reduce link switching delay; the presented technique also prevents unnecessary link switching [7]. Thai-Chien Bui et al [8] presented a flexible link switching algorithm based on the received power by considering adaptive hysteresis margin.

In this paper, we propose a novel position prediction link switching algorithm for VLC configuration. Part II of the paper describes the background and proposed System model. Part III presents the link switching algorithm developed. Part IV deals with the results and finally conclusions are drawn in part V.

## II. BACKGROUND & PROPOSED SYSTEM MODEL

The system comprises empty conference room of dimensions 12m x 12m and height 3m, with 9 white LED - based lighting cell array (3x3) mounted on the room ceiling with LED coverage. The VLC indoor system adequately fulfills the illumination requirement since illumination is the primary function of LED lighting system. Further the system can support the device mobility with continuous data transmission.

This LED distribution provides the illumination level specified in ISO standard (illumination 300 ~1500 lx in all the positions in the room) [9]. Thus, nine Access points are in positions (-5; -5;1.2), (-5;5;1.2), (5; -5;1.2), (5;5;1.2), (-5;0;1.2), (5;0;1.2), (0; -5;1.2), (0;5;1.2), and (0;0;1.2) as shown in fig 1. Movement of the mobile receiver device is assumed to be in a plane that is situated 1.8 m below the LED.

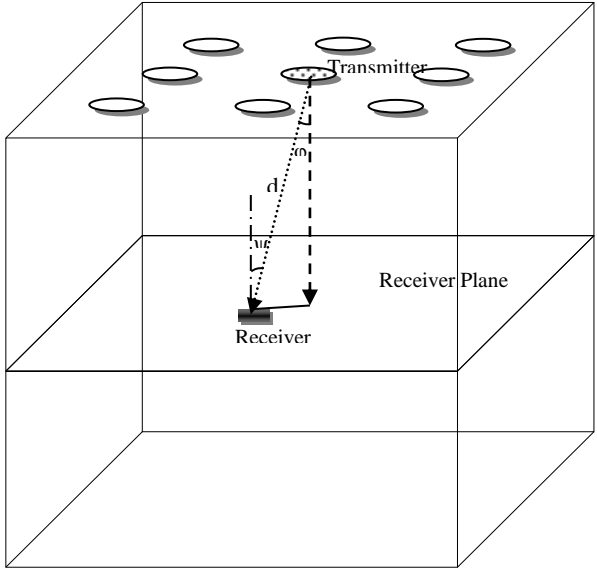

Fig 1: 3D scenario with 9 LEDS and a moving receiver

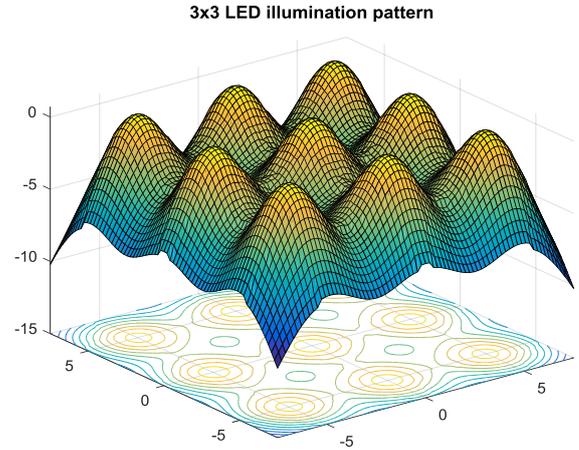

Fig. 2: 3D Plot for Received Power Distribution

Further, each LED source produces a circular lighting field. Thus, each lighting cell has omnidirectional signal coverage with radius r meter and each LED Access Point (AP) transmits at data rate D bits/sec at the center of the cell and a MT receives data when it is inside a lighting cell. Given a mobile terminal crossing a lighting cell at constant speed v [m/s], the cell gain [bits] is the number of bits received by the mobile terminal during the time interval it spends while traversing crossing the lighting cell.

The parameters that need to be monitored for performance assessment of the VLC system are the transmitted optical power and luminous intensity. Horizontal luminous intensity is given by (1) at the receiver plane when distance between the detector surface and LED is d

$$I_{hor} = I(0) cos^{m_1}(\varphi) I d^2 . cos(\psi) \quad (1)$$

Where, $\varphi$ is irradiance angle with respect to the axis, normal to the transmitter surface and $\psi$ is the incidence angle with respect to the axis, normal to the receiver plane. I (0) is the luminous intensity at the middle point just below the light, and $m_1$ is Lambertian emission order. If $T_s(\psi)$ is the filter transmission function, received power is

$$P_r = P_t \frac{(m_1 + 1)}{2\pi d^2} cos^{m_1}(\varphi) . T_s(\psi) . g(\psi) . cos(\psi) \quad (2)$$

A simple linear optical AWGN channel is considered with the optical transmitted power $P_t(t)$, impulse response $h(t)$ and signal independent additive noise $n(t)$ [10].
Thus, VLC channel can be modeled as:

$$I_p(t) = RP_t \otimes h(t) + n(t) \quad (3)$$

With transmitted optical power

$$P_t = \lim_{T \to \infty} \frac{1}{T} \int_0^T P_i(t) dt \qquad P_i > 0 \quad (4)$$

Indoor scenario also requires consideration of reflections from the walls. Thus, total Received power is given by 5 [12]

$$P_r = \sum \{P_t H_d(0) + \int P_t dH_{ref}(0)\} \quad (5)$$

Where $H_d(0)$ directed path and first reflection at the receiver is:

$$H_{ref} = \rho d_{LED} A_{wall} cos^{m_1}(\varphi_r) \cos(\alpha_{ir}) \cos(\beta_{ir}) T_s(\psi) g(\psi) \cos(\psi_r) \quad (6)$$

Where, $d_{LED}$ is the distance between the LEDs, $\rho$ is reflectance factor, $A_{wall}$ is reflective area of infinitely small region, $\alpha$ and $\beta$ are the angle of irradiance to a reflective point and $\psi_r$ is the angle of incidence at the reflective surface. 3 D plot of the received optical power at different points in the room is plotted in Fig.2

### III. PROPOSED LINK SWITCHING MECHANISM

Link switching techniques may be broadly categorized as:
a) Hard Link Switching
b) Soft Link Switching

Hard link switching is the conventional way to perform the switching process. When strength of received signal goes down the threshold level, switching process is initiated. First user scans the neighborhood to find the best transmitter and subsequently sends the request to coordinator to switch the link; which accepts the request and sends instructions to disconnect the current link. Once, this link is disconnected, next link association request is generated by the user device. Coordinator checks with the designated transmitter and if response of this transmitter is positive, the link association is performed. Main drawback of hard link switching is disconnection from the current transmitter before association from the next transmitter. This results in service disruption and QoS degradation. Further scanning period increases the switching delay in addition to the delay involved in switching process. Whereas Soft link switching ensures that the user device is disconnected from the current serving transmitter only when link between the new serving transmitter and user device is established. This procedure ensures that there is no interruption in service and QoS is maintained. But for this we must pay a price for complex and expensive hardware, and user device is needed to receive signals from two or more transmitters simultaneously thus waste resources, increases data traffic and may result in downlink interference. Also, there is possibility of service interruption or low QoS , as sometimes , user device may not receive signal due to interference. In some cases , selected transmitter may not have enough resources to accommodate the user device.

Presently, methods available for link switching are:
Each UD(User device) sends back acknowledgement to the coordinator whenever it receives data. if coordinator does not receive acknowledgement in predetermined time, it searches for the acknowledgement in the same time slot in adjacent cells and if it found in other cell, control is transferred to corresponding cell.
Pre scan & RSS : In idle time UD checks signal from neighboring cells, makes a table decides to switch link based on three thresholds.

We are hereby proposing a hard link switching algorithm, which does not require any complex circuitry and follows 802.15.7 standards for visible light communication. Our Link switching process consists of following procedures:
- Position estimation of mobile user
- Recognition of the correct transmitter
- Handover

First, we consider that in an indoor scenario where user devices are generally hand held devices and the speed of the owner of the device i.e. user is not much. We can also take the liberty that user generally moves approximately in a single direction with time and does not roam around arbitrarily. Second, we assume that with fixed interiors and light fixtures, the coordinator already has the information about the best suitable transmitter for any arbitrary position. With all these assumptions and information we propose a simple method of link switching which involves estimation of the future position of the user device and switching of transmitter links by the time user reaches that position. When user device receives or sends data in between two active periods there is an inactive or idle period. During this period UD scans the scenario and measures the strength of all the signals it receives. Then UD sends this measurement report to the coordinator. So, the proposed process starts with estimation of position of user.

*A. Position Estimation :*

In this paper, triangulation method has been used to determine the location of receiver since it provides better accuracy compared to global positioning system (GPS) in indoor environment. [12]. This technique helps to determine the location of receiver based on known positions of three transmitters. If $T_1$, $T_2$, $T_3$ are three transmitters with known positions as shown in Fig. 3, then the position of receiver Rx (x, y) may be determined by solving the following equations where $d_1$, $d_2$, $d_3$ are derived based on the RSS from corresponding Txs.

$$(x - x_1)^2 + (y - y_1)^2 = d_1^2 \quad (7)$$

$$(x - x_2)^2 + (y - y_2)^2 = d_2^2 \quad (8)$$

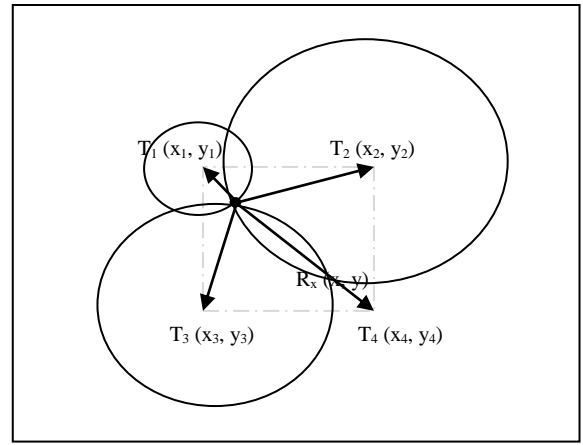

Fig 3: Triangulation method for prediction of Rx point

$$(x - x_3)^2 + (y - y_3)^2 = d_3^2 \quad (9)$$

Coordinator bounds the channel time in superframe structure. A superframe as shown in Fig. 4 is started and ended by the transmission of a beacon frame and can have an active portion as well as an inactive portion. During each inactive portion, user device measures the Received Signal Strength (RSS) of all the nearby transmitters and sends the measurement report repeatedly to the coordinator. Coordinator uses these RSS values to determine the current position of the user using triangulation method. Call flow diagram of our proposed link switching algorithm is shown in Fig. 5.

Coordinator thus generates a path report for every UD and adds its new position after every $dt_1$ time interval (time between two inactive portions). Coordinator can predict the next position of the user using its previous path. If UD was at position $1(x_1, y_1)$ at time $t = 0$ and at position 2 at time t = dt$_1$, where dt$_1$ is the time interval between two inactive portions i.e. time interval between two measurement reports send by the user. Coordinator would calculate the next position of the UD at time t = 2dt$_1$; time when user sends next measurement report.

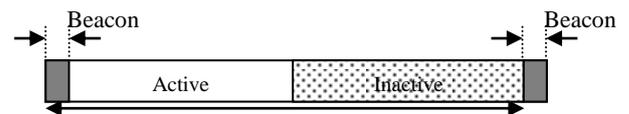

Fig. 4: Superframe Structure[2]

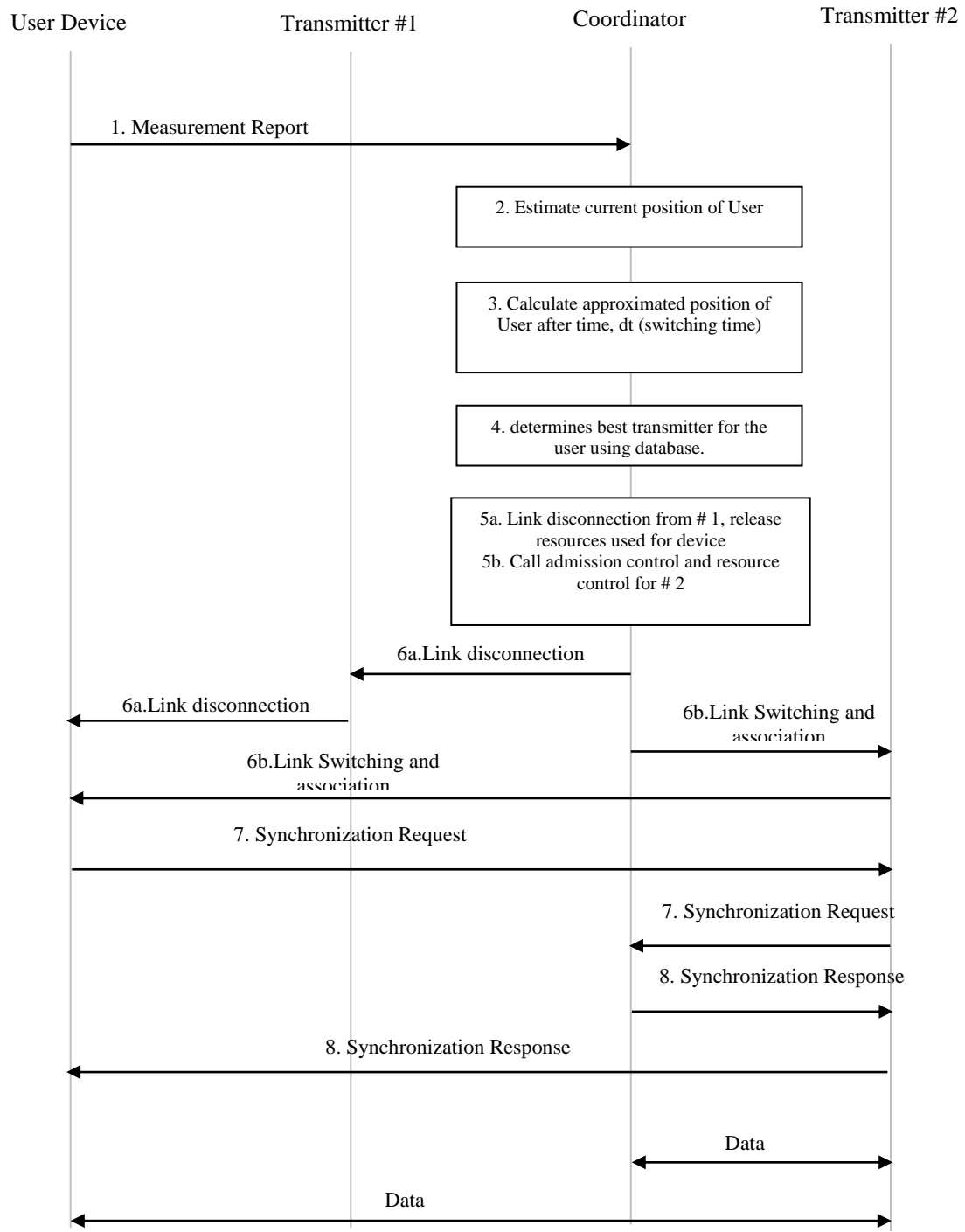

Fig 5: Call flow Diagram for proposed link switching algorithm

$$x_e = x_2 + \frac{(x_2 - x_1)}{2} \quad (10)$$

$$y_e = y_2 + \frac{(y_2 - y_1)}{2} \quad (11)$$

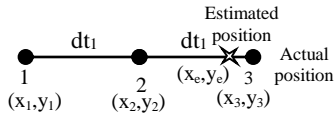

Fig 6: Current position estimation

Since user is moving with approximately same speed and direction; estimated position (xe, ye) will obviously be close enough to the actual position (x3, y3) (Fig. 6).
Coordinator uses this information to find the most suitable transmitter for this future position (Fig. 7). If this transmitter is different from the current serving transmitter; link switching process is initiated. Coordinator initiates both the disconnection process from the current transmitter and link association process with next transmitter simultaneously. This reduces the switching time and ensures minimum service disruption. By the time user reaches the next position, next transmitter starts the service. If dt$_2$ is the time taken by the link switching process and dt$_1$ is the time to reach next known position

Then

$$dt_1 \approx dt_2 \quad (12)$$

if dt$_1$ is nearly equal to dt$_2$ successful handover is done and user enjoys uninterrupted service.
Traditional link switching delay can be calculated as:

$$T_{trad} = t_{scan} + t_{decision} + t_{discon} + t_{linksw} + t_{linkasso} + t_{sync}; \quad (13)$$

Where tscan is the time to scan the neighboring transmitter after RSS reduced to a given threshold level, t$_{decision}$ is time taken by the user device to select the next transmitter, t$_{discon}$, t$_{linksw}$, t$_{linkasso}$ and t$_{sync}$ are respectively time taken in present link disconnection process, new link establishment request, link association and synchronization process. Use of position prediction technique efficiently improves link switching delay time by eliminating the need of scanning and decision making time by user device, in addition, link disconnection and new link establishment processes are performed simultaneously. This results in reduced link switching delay time. Link switching delay time in proposed scheme is given by (14)

$$T_{pred} = t_{discon/linksw} + t_{linkasso} + t_{sync}; \quad (14)$$

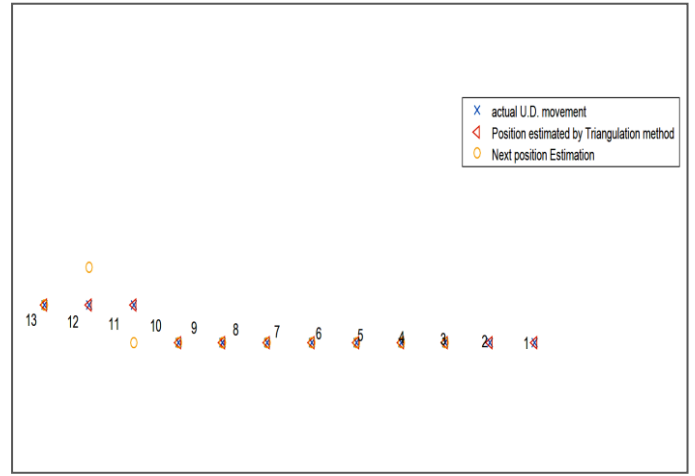

Fig 7: Current position estimation , Actual UD position Future position Prediction of MT

IV. CONCLUSION

The proposed scheme relies on the pre-scanned information of the indoor scenario stored in the coordinator's database; which can be refreshed at regular intervals. This Link switching method is based on the consideration that indoor scenario is static with reference to the position of the objects and LEDs . Any changes, if occur, can be incorporated in the coordinator database. Since, link switching is done automatically so time delay due to link disconnection request and link switching is reduced. Thus, overall switching time is also reduced. Further, user need not initiate the link disconnection request and link switching request and thus switching time is reduced. Predicted position accuracy can be increased further by considering three or more previous positions; thus reducing link switching error. If link switching failure occurs three or more time at a position, coordinator database is updated for the same. In addition, this link switching method does not require any complex circuitry or any software changes in User Device and ensures high QoS.